\documentclass[%
 reprint,
 amsmath,amssymb,
 aps,
]{revtex4-2}

\usepackage{graphicx}
\usepackage{graphicx, xurl, hyperref}
\usepackage{color}
\usepackage{float}
\usepackage{dcolumn}
\usepackage{bm}
\begin{document}

\preprint{APS/123-QED}

\title{Deep Neural Networks can eliminate Spiral-wave Turbulence in Cardiac Tissue Models}

\author{Vasanth Kumar Babu}
	\email[]{vasanthb@iisc.ac.in}
		\affiliation{Centre for Condensed Matter Theory, Department of Physics, Indian Institute of Science, Bangalore, 560012, India. }
	\author{Rahul Pandit}
		\email[]{rahul@iisc.ac.in}
	\affiliation{Centre for Condensed Matter Theory, Department of Physics, Indian Institute of Science, Bangalore, 560012, India. }
 
\date{\today}

\begin{abstract}

Ventricular arrhythmias, like ventricular tachycardia (VT) and ventricular fibrillation (VF), precipitate sudden cardiac death (SCD), which is the leading cause of mortality in the industrialised world. Thus, the elimination of VT and VF is a problem of paramount importance, which is studied experimentally, theoretically, and numerically. Numerical studies use partial-differential-equation models, for cardiac tissue, which admit solutions with spiral- or broken-spiral-wave solutions that are the mathematical counterparts of VT and VF. \textit{In silico} investigations of such mathematical models of cardiac tissue allow us not only to explore the properties of such spiral-wave turbulence, but also to develop mathematical analogues of low-amplitude defibrillation by the application of currents that can eliminate spiral waves. We develop an efficient deep-neural-network U-Net-based method for the control of spiral-wave turbulence in mathematical models of cardiac tissue. Specifically, we use the simple, two-variable Aliev-Panfilov and the ionically realistic TP06 mathematical models to show that the lower the correlation length $\xi$ for spiral-turbulence patterns, the easier it is to eliminate them by the application of control currents on a mesh electrode. We then use spiral-turbulence patterns from the TP06 model to train a U-Net to predict the sodium current, which is most prominent along thin lines that track the propagating front of a spiral wave. We apply currents, in the vicinities of the predicted sodium-current lines to eliminate spiral waves efficiently. The amplitudes of these currents are adjusted automatically, so that they are small when $\xi$ is large and \textit{vice versa}. We show that our U-Net-aided elimination of spiral-wave turbulence is superior to earlier methods.

\end{abstract}

\maketitle


\section{Introduction\label{sec:Intro}} 

Sudden cardiac death (SCD), primarily induced by ventricular arrhythmias, remains a major cause of mortality worldwide~\cite{mehra2007global,nowbar2019mortality,khairy2022sudden}. Unfortunately, SCDs continue to increase, across all population segments, and they affect a significant fraction of young people and athletes~\cite{stormholt2021symptoms,roshan2023multiscaleI,roshan2023multiscaleII}. The annual global death toll because of cardiovascular diseases is projected to rise~\cite{angell2020american} from $\simeq 17.9$ million in 2019 to over $23.6$ million by 2030. To address the challenges posed by these life-threatening arrhythmias, it is imperative to investigate them experimentally, theoretically, and numerically. 

Numerical studies are playing an increasingly important role in these investigations. Such numerical investigations employ detailed partial-differential-equation models for cardiac tissue that account for (a) the transmembrane potential in ventricular myocytes, (b) the ionic currents whose dynamics is governed by ion channels and the gating variables that are associated with these channels, and (c) the spatiotemporal evolution of the resulting waves of electrical activation that propagate through cardiac tissue by virtue of the gap-junction couplings between the myocytes. Such models can exhibit spiral (or scroll) waves of electrical activation; single spiral waves are associated with ventricular tachycardia (VT) whereas broken-spiral-waves, or spiral-wave turbulence, are associated with ventricular fibrillation (VF). Clinically, VT is often a precursor to VF; the latter leads to uncontrolled quivering of the ventricles, so the left ventricle is unable to pump oxygenated blood to all parts of the body, and in the absence of medical intervention, sudden cardiac death occurs in about two-and-a-half minutes. 

The elimination of VT and VF is attempted by both pharmacological and electrical means. The latter are especially useful in emergencies. Electrical defibrillation is carried out typically using external defibrillators~\cite{jonsson2023automated} or, in high-risk patients, implantable cardioverter-defibrillators (ICDs)~\cite{maron2023development}. However, even when such defibrillation succeeds in removing spiral or scroll waves, it uses a large amount of energy that can potentially cause significant tissue damage and result in additional complications~\cite{tereshchenko2009transient,nusair2010electric,mackenzie2004making,moss2012reduction}. Therefore, the optimisation of such defibrillation schemes and the development of low-energy applications are of utmost importance~\cite{madhavan2013optimal}. It is also a grand challenge for \textit{in vivo}, \textit{ex vivo}, and \textit{in vitro} experiments [see, e.g., Refs.~\cite{gray1995mechanisms,pandit2013rotors,luther2011low,jenkins2022inspection}] and for \textit{in silico} studies of the control of spiral-wave turbulence, in mathematical models for cardiac tissue, whose importance is increasing in the study and control of mathematical analogues of VT and VF~\cite{cherry2008visualization,aliev1996simple,sinha2001defibrillation,ten2004model,ten2006alternans,sridhar2008controlling,alcaraz2010review,shajahan2009spiral,mulimani2020deep,detal2022terminating,luo2024unpinning,mulimani2023overview}. 

Spiral-wave turbulence, which displays multiple, fragmented self-sustaining spiral waves, is a form of spatiotemporal chaos so its characterization and control is very difficult. A quantification of the spatial organization of such turbulence can be used to mitigate this chaos to some extent~\cite{bayly1998spatial,alonso2004expanding,reichenbach2007noise,maselko1991chemical}. Indeed, both experiments and numerical simulations indicate that, in a state with spiral-wave turbulence, the spatial correlation function of the transmembrane potential $V_m$ decays rapidly~\cite{bayly1998spatial,alonso2004expanding}. Furthermore, experimental studies~\cite{hsia1996defibrillation,barbaro2001effect,everett2001assessment,everett2001frequency,alcaraz2010review} suggest that the success of defibrillation depends on the correlation length of this decaying correlation function. 

We use this crucial insight to design a new and effective machine-learning-aided defibrillation scheme, which specifically accounts for the spatial organization of electrical waves in spiral-turbulence patterns. We first use two mathematical models for cardiac tissue, namely, the Aliev-Panfilov model~\cite{aliev1996simple}, a simple two-variable model, and the Ten Tusscher-Panfilov (TP06) model~\cite{ten2004model,ten2006alternans}, an ionically realistic human-ventricular-myocyte model, based on experimental data. We demonstrate that, for both these models, the mesh-defibrillation scheme of Refs.~\cite{sinha2001defibrillation,shajahan2009spiral} eliminates spiral-excitation patterns more easily if the correlation 
length $\xi$ of the spatial correlation function is large, and there are few phase singularities (spiral-wave cores), than if $\xi$ is low, when the number of singularities increases. Reference~\cite{mulimani2020deep} has developed a deep-learning method that identifies spiral waves and their cores, and then generalises the mesh-based defibrillation method~~\cite{sinha2001defibrillation,shajahan2009spiral} by applying control currents in the vicinities of these cores. 

We develop a U-Net-based defibrillation approach that goes well beyond the defibrillation schemes of Refs.~\cite{sinha2001defibrillation,shajahan2009spiral,mulimani2020deep}. Our new scheme uses data from the realistic TP06 model to train a U-Net to \textit{predict} the sodium current, which is significant along fine lines that track the propagating arm of a spiral wave of electrical activation. This leads us to the following natural and optimal defibrillation strategy: We apply control currents in the vicinities of the fine lines associated with the sodium current predicted by the U-Net; in this scheme, the total strength of the applied current decreases automatically with an increase in $\xi$.




\section{Results}
\subsection{Mesh-defibrillation Scheme\label{sec:results}}

We first employ the mesh-defibrillation scheme of Ref.~\cite{sinha2001defibrillation} to illustrate the elimination of spiral-turbulence in 
the Aliev-Panfilov [Section~\ref{subsec:ReAP}] and TP06 [Section~\ref{subsec:ReTP06}] models. We show that the larger the spatial correlation length of a
spiral-turbulence pattern, the more successful is the mesh defibrillation scheme in eliminating this turbulence. We then uncover the relation between the phase singularities, spiral arms, and this correlation length in Section~\ref{subsec:phsing}. In Section~\ref{subsec:U-Netdefib}, we present our main results, which we obtain by using deep-learning methods to develop a new defibrillation scheme; our new scheme applies currents to the regions where the sodium current peaks, along the spiral arms; this allows us to control the total applied current in a manner that depends on the correlation length.

 \begin{figure*}
    \centering
    \includegraphics[width=\textwidth]{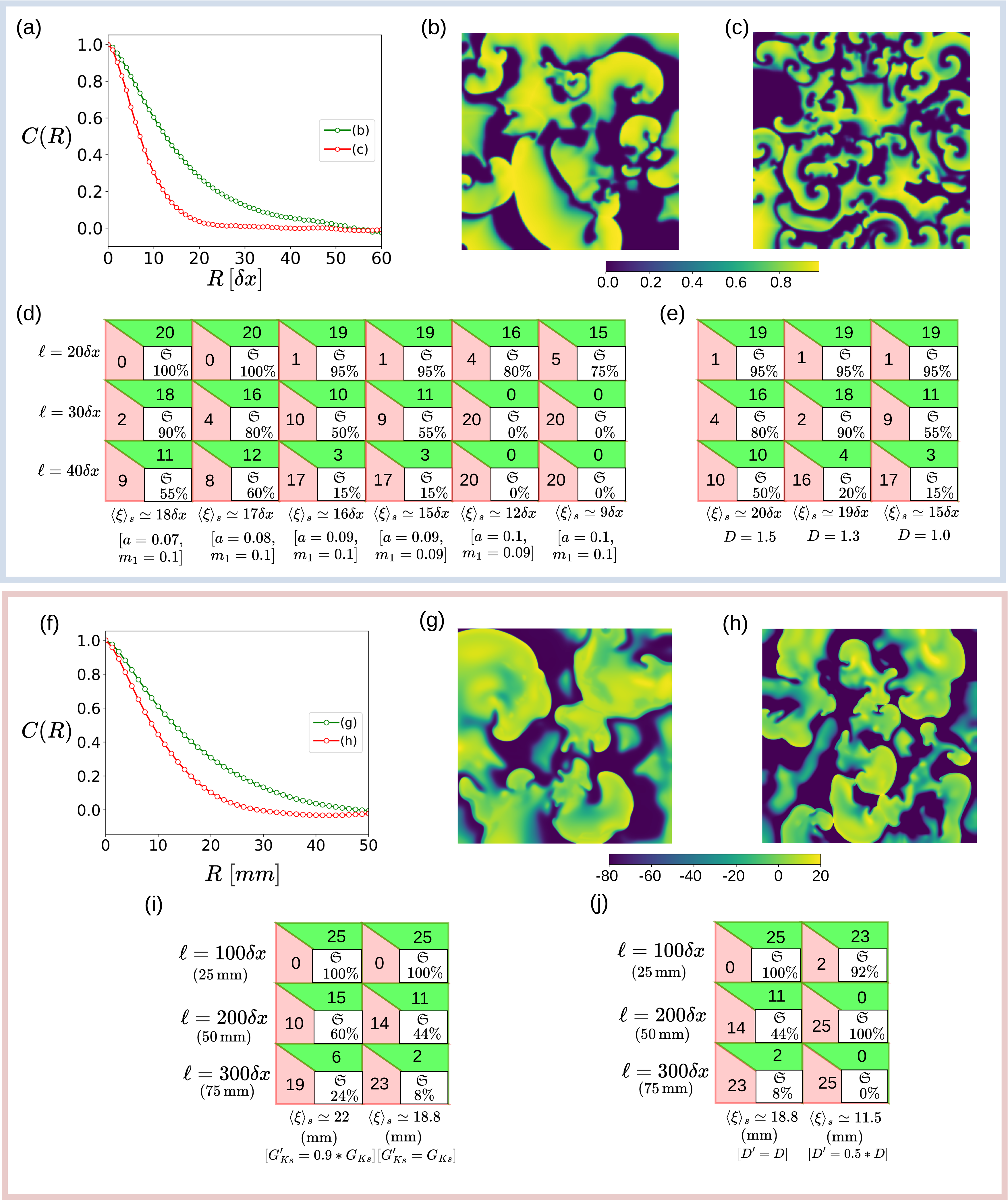}
    \caption{(a) Plot of the spatial correlation function $C(R)$ [Eq.~\eqref{eq:corr}] versus the separation $R$ [in grid points ($\delta x$)] in red and green, for two illustrative spiral-turbulence patterns in the Aliev-Panfilov model~\eqref{eq:AP}, shown via pseudocolor plots of $u$, in (b) and (c), respectively; these are obtained from our numerical solution of Eq.~\eqref{eq:AP}. (d) The upper green quadrilateral (lower red quadrilateral) gives the number of successful (unsuccessful) cases of the elimination of spiral-turbulence patterns, via mesh defibrillation [see Section~\ref{subsubsec:MD}] for $20$ spiral-turbulence patterns for each set of paramters, $a$ and $m_{1}$ [with all other parameters fixed in Eq.~\eqref{eq:AP}], which are given at the bottom, along with the mean correlation length $\langle\xi\rangle_{s}$ [see text]; $\mathfrak{S}$ is the elimination success rate.  (f), (g), (h), (i) and (j) are the TP06 model~\eqref{eq:TP06} counterparts of (a), (b), (c), (d), and (e), respectively [see text]; for this model the colorbars for our pseudocolor plots cover the range $[-80, 20]$ for the transmembrane potential.}
    \label{fig:AV_TP06_tableST}
\end{figure*}


 \begin{figure*}
    \centering
    \includegraphics[width=\textwidth]{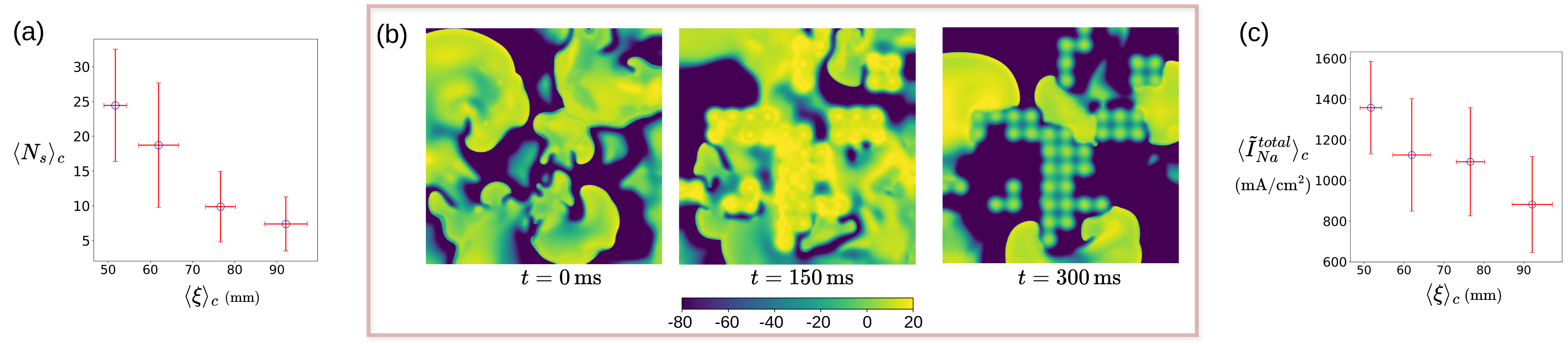}
    \caption{(a) Mean number of phase singularities~\eqref{eq:Ph_In} $\langle N_s \rangle_c$ versus the mean correlation length $\langle\xi\rangle_{c}$ for the TP06 model~\eqref{eq:TP06}. (b) Illustration of the application of an external current on the phase singularities in the TP06 model~\eqref{eq:TP06}. (c) The mean of the total normalized sodium current $\langle\tilde{I}_{Na}^{total}\rangle_{c}$~\eqref{eq:Itot} versus $\langle\xi\rangle_{c}$.}
    \label{fig:phasesingres}
\end{figure*}

 \begin{figure*}
    \centering  \includegraphics[width=0.9\textwidth]{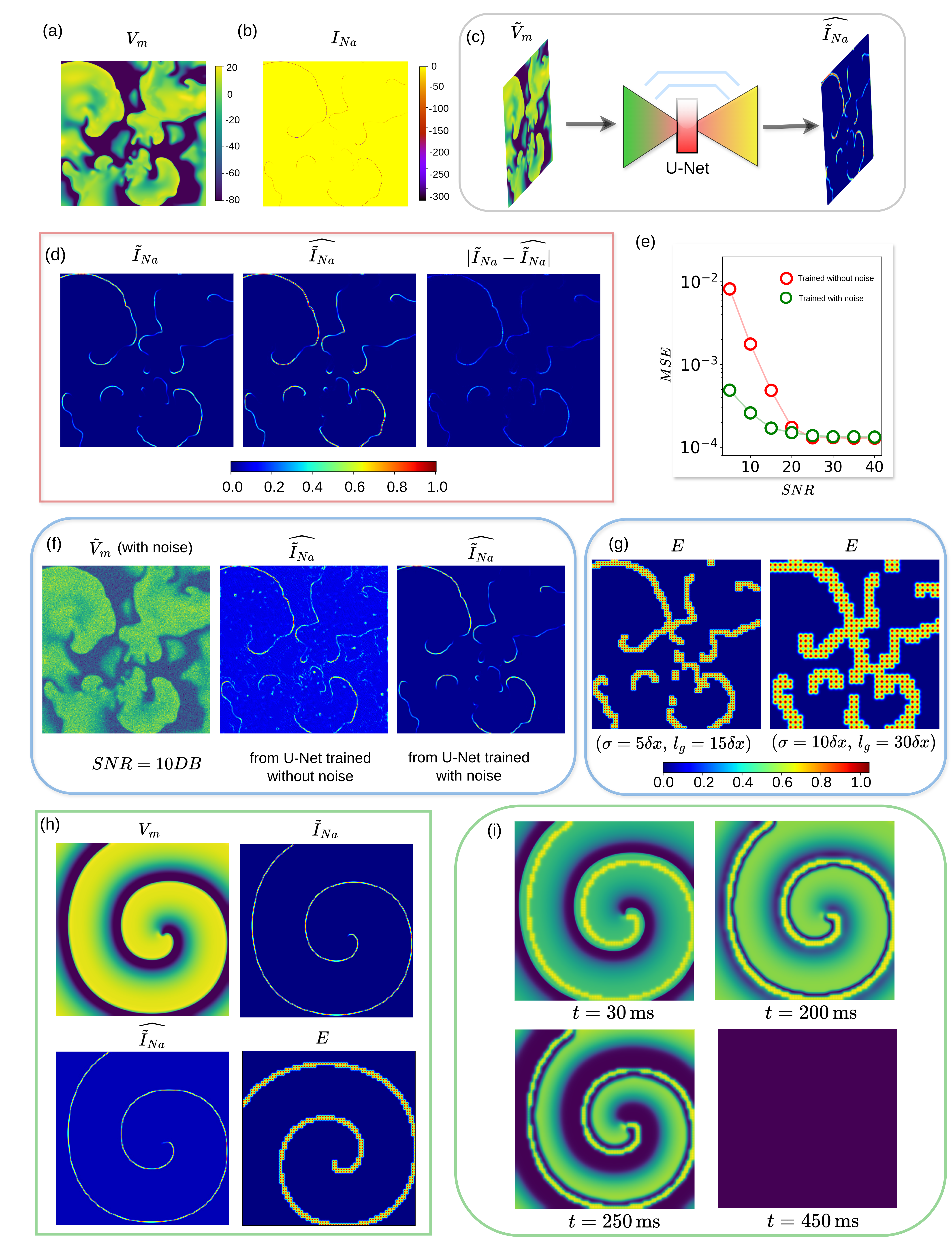}
    \caption{
    (a) and (b) Illustrative pseudo-color plots of $V_{m}$ and the corresponding $I_{Na}$. (c) A schematic diagram of the mapping between the normalized $V_{m}$ [$\tilde{V}_{m}$] and the normalized $I_{Na}$ [$\tilde{I}_{Na}$] using the U-Net. (d) Illustrative pseudo-color plots of $\tilde{I}_{Na}$, the corresponding U-Net prediction $\widehat{\tilde{I}_{Na}}$, and $|\widehat{\tilde{I}_{Na}}-\tilde{I}_{Na}|$. (e) Mean-squared-error (MSE) loss for the validation data with various values of the signal-to-noise ratio
(SNR) for the U-Net, trained
with and without noise in the input $\tilde{V}_{m}$. (f) An illustrative pseudocolor plot of
$\tilde{V}_m$ , with SNR$=10$ decibel (dB), and the corresponding
predictions $\widehat{\tilde{I}_{Na}}$ from our U-Net, trained both with and
without noise. (g) Illustration of electrodes that are switched
on in the vicinity of $I_{Na}$ activation [see text]. (h) Pseudocolor plots of a spiral-wave [${V}_{m}$], the associated $\tilde{I}_{Na}$, the U-Net prediction $\widehat{\tilde{I}_{Na}}$, and $E$, the electrodes activated in the vicinity of $\widehat{\tilde{I}_{Na}}$. (i) Illustration of the elimination of the spiral wave by the application
of a current on the electrodes $E$ [$t=0$ in (h)].
    }
    \label{fig:INa_pred}
\end{figure*}

 \begin{figure*}
    \centering
    \includegraphics[width=\textwidth]{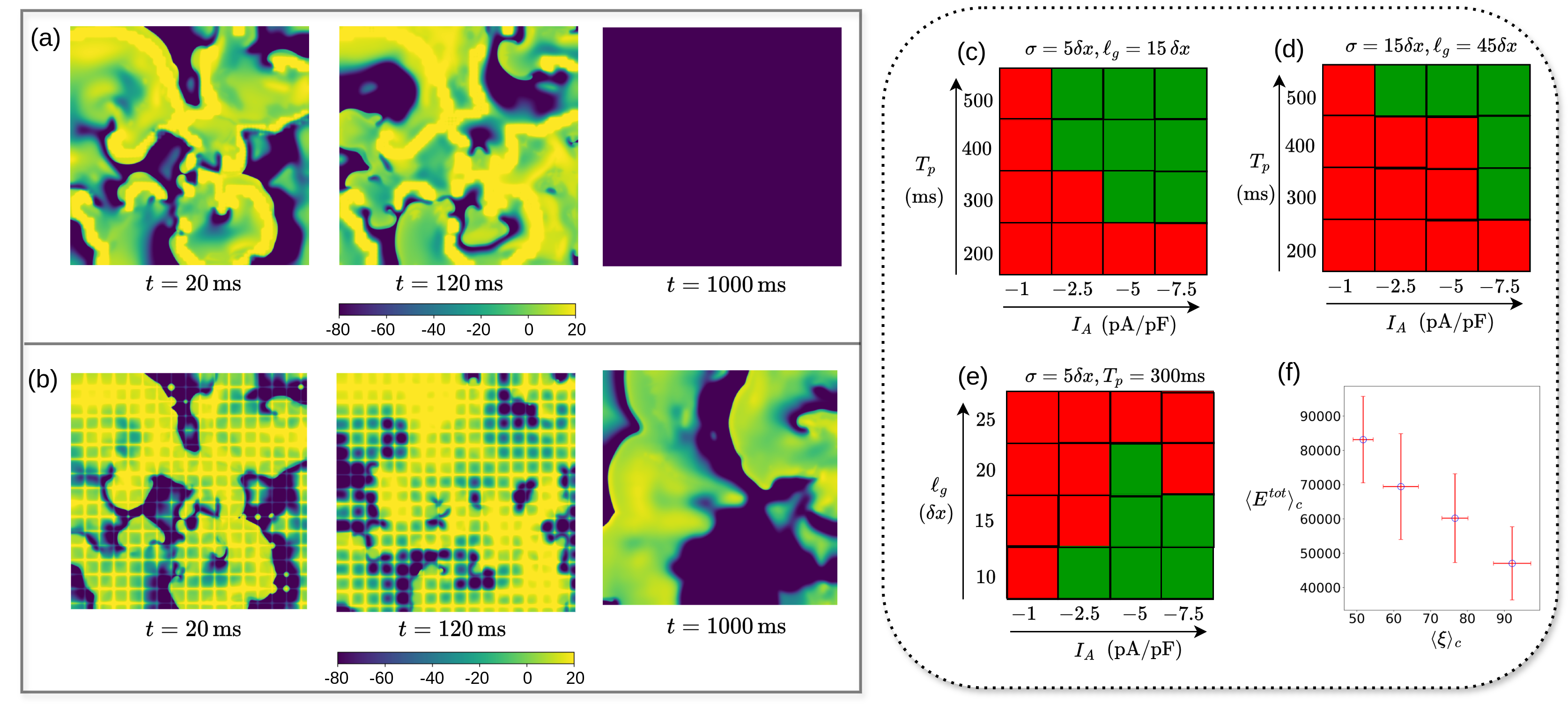}
     \caption{(a) Pseudocolor plots of the transmembrane potential $V_m$ illustrating the elimination of spiral-turbulence patterns in the TP06~\eqref{eq:TP06_1} model, where we apply currents in the vicinity of the peaks of the sodium currents predicted by the U-Net [$t=0$ in Fig.~\ref{fig:INa_pred} (a)]; (b) the counterparts of (a) for the mesh-defibrillation scheme (see text). Panels (c), (d), and (e) show the success of our U-Net-based defibrillation scheme for various choices of the parameters $I_A$, $T_p$, $\sigma$, and $\ell_{g}$ in Eqs.~\eqref{eq:Kernal} and \eqref{eq:Ele}; green and red squares indicate successful and unsuccessful outcomes, respectively. (f) $\langle E^{total} \rangle_{c}$ [proportional to the mean of the total applied current] versus the mean correlation length $\langle \xi \rangle_{c}$.}
    \label{fig:defib_result}
\end{figure*}
\subsubsection{Aliev-Panfilov Model}\label{subsec:ReAP}

In Figs.~\ref{fig:AV_TP06_tableST} (a)-(e), we summarize the outcomes of the mesh-defibrillation scheme [Section ~\ref{subsubsec:MD}] on the elimination of spiral-wave turbulence in the Aliev-Panfilov model~\eqref{eq:AP}. Our simulations for this model produce spiral-turbulence patterns with a spatial-correlation length $\xi$ [see Section~\ref{subsubsec:SPF} for details], which depends on the model parameters [see Section~\ref{subsec:AP}]: In Fig.~\ref{fig:AV_TP06_tableST} (a), we plot the spatial correlation function $C(R)$ [Eq.~\eqref{eq:corr}] versus the separation $R$ in red and green, for two illustrative spiral-turbulence patterns, shown via pseudocolor plots of $u$, in Figs.~\ref{fig:AV_TP06_tableST} (b) and (c), respectively; these are obtained from our numerical solution of Eq.~\eqref{eq:AP}.

The upper green quadrilateral (lower red quadrilateral) of the boxes in the Table in Fig.~\ref{fig:AV_TP06_tableST} (d) gives the number of successful (unsuccessful) cases of the elimination of spiral-turbulence patterns, via mesh defibrillation~\ref{subsubsec:MD}. We use a total of $20$ such patterns [see Appendix~\ref{subsec:sim_det}], for each set of parameters, $a$ and  $m_{1}$ [with all other parameters fixed in Eq.~\eqref{eq:AP}], which are given at the bottom of this Table along with the mean correlation length ${\langle\xi\rangle}_{s}$ [averaged over these $20$ patterns];  ${\langle\xi\rangle}_{s}$ decreases from left to right. Each row of this Table is labelled by $\ell$, the mesh spacing [Fig.~\ref{fig:mesh}]; the columns are labelled by $a$, $m_{1}$, and  $\langle\xi\rangle_{s}$; for each box here, the current $I_{ext}=-0.5$~\eqref{eq:AP} and its application time $T_{p} = 100\,(5000\delta t)$ [see Section~\ref{subsubsec:MD}].
If we consider the first and last boxes in the second row with $\ell=30\delta x$, 
we find: (i) $\langle\xi\rangle_{s}\simeq18\delta x$ and an elimination-success rate $\mathfrak{S} = 90\%$;
and (ii) $\langle\xi\rangle_{s}\simeq9 \delta x$ and $\mathfrak{S} = 0\%$. Similar results hold for $\ell=20\delta x$ (row 1) and $\ell=40\delta x$ (row 3) and also for different values of $I_{ext}$ and $T_{p}$ [see Appendix~\ref{app:Add_mesh_details}].

In the Table in Fig.~\ref{fig:AV_TP06_tableST} (e), we vary $D$ [keeping other parameters fixed] in Eq.~\eqref{eq:AP},
so the columns are now labelled by the value of $D$ and the computed value of $\langle\xi\rangle_{s}$, which decreases from left to right [again we use  $20$ spiral-turbulence patterns]. For each box here, $I_{ext}=-0.5$ and $T_p=100\,(5000\delta t)$. If we consider the first and last boxes in the second row with $\ell=30\delta x$, we find: (i) $\langle\xi \rangle_{s}\simeq19\delta x$ and $\mathfrak{S}=80\%$ (ii) $\langle\xi\rangle_{s}\simeq13\delta x$ and $\mathfrak{S}=30\%$. Similar results hold for $\ell=20\delta x$ (row 1) and $\ell=40\delta x$ (row 3) and also for different values of $I_{ext}$ and $T_{p}$ [see Appendix~\ref{app:Add_mesh_details}].

\subsubsection{TP06 Model}\label{subsec:ReTP06}
We now consider mesh defibrillation for the TP06 model~\eqref{eq:TP06}. 
Figures~\ref{fig:AV_TP06_tableST} (f), (g), (h), (i), and (j) are the TP06 counterparts of Figs.~\ref{fig:AV_TP06_tableST} (a), (b), (c), (d), and (e), respectively.
In Fig.~\ref{fig:AV_TP06_tableST} (f), we plot $C(R)$ [Eq.~\eqref{eq:corr}] versus the separation $R$ in red and green, for two illustrative spiral-turbulence patterns, shown via pseudocolor plots of $V_m$, in Figs.~\ref{fig:AV_TP06_tableST} (g) and (h), respectively; these are obtained from our numerical solution of Eq.~\eqref{eq:TP06}.

The rows in Table in Fig.~\ref{fig:AV_TP06_tableST} (i) are labelled by $\ell$; the columns are labelled by the conductance $G_{Ks}$, which we vary [keeping other parameters fixed in Eq.~\eqref{eq:IKs}] and $\langle\xi\rangle_{s}$, which we obtain by averaging over $25$ spiral-turbulence configurations [see Appendix~\ref{subsec:sim_det}];
$\langle\xi\rangle_{s}$ increases from left to right. 
For each box in this Table, the current $I_{ext}=-100\,\frac{\mathrm{pA}}{\mathrm{pF}}$~\eqref{eq:TP06} and its application time $T_{p} = 15000\delta t\, (300\,\mathrm{ms})$. 
If we consider the first and second boxes in the second row with $\ell=200\delta x\,(50\,\mathrm{mm})$, we find: (i) $\mathfrak{S}=60\%$ for $\langle\xi\rangle_{s}\simeq88\delta x\,(22\,\mathrm{mm})$; and (ii) $\langle\xi\rangle_{s}\simeq75\delta x (18.8\,\mathrm{mm})$, $\mathfrak{S}=44\%$. Similar results hold for $\ell = 25\,\mathrm{mm}$ (row 1) and $\ell = 75\,\mathrm{mm}$ (row 3) and also for different values of $I_{ext}$ and $T_{p}$ [see Appendix~\ref{app:Add_mesh_details}].

 In Table in Fig.~\ref{fig:AV_TP06_tableST} (j), we vary $D$ [keeping other parameters fixed in Eq.~\eqref{eq:TP06}], which labels the columns along with $\langle\xi\rangle_{s}$. 
 If we now consider the first and second boxes in the second row with $\ell=200\delta x\;(50\,\mathrm{mm})$, we find: (i) $\mathfrak{S}=44\%$ for $\langle\xi\rangle_{s}\simeq75\delta x \;(\simeq 18.8\,\mathrm{mm})$; and (ii) $\langle\xi\rangle_{s} \simeq46\delta x \;(\simeq11.5\,\mathrm{mm})$, $\mathfrak{S}=0\%$. Similar results hold for $\ell= 25\,\mathrm{mm}$ (row 1) and $\ell = 75\,\mathrm{mm}$ (row 3) and also for different values of $I_{ext}$ and $T_{p}$ [see Appendix~\ref{app:Add_mesh_details}].

\subsection{Phase singularities, spiral arms and the correlation length}\label{subsec:phsing}

From our discussion of the mesh-defibrillation scheme in Sections~\ref{subsec:ReAP} and ~\ref{subsec:ReTP06}],
and Fig.~\ref{fig:AV_TP06_tableST}, 
we see that, in both the Aliev-Panfilov~\eqref{eq:AP} and TP06~\eqref{eq:TP06} models, the higher the spatial correlation length $\langle\xi\rangle_{s}$,
the larger is $\mathfrak{S}$, i.e., the more effective is the elimination of spiral-turbulence patterns.
Our results here are consistent with those obtained from experimental studies~\cite{hsia1996defibrillation,barbaro2001effect,everett2001assessment,everett2001frequency,alcaraz2010review}.
Henceforth, we concentrate on the TP06 model.

We expect that $\langle\xi\rangle_{c}$ (and, concomitantly, $\mathfrak{S}$) decreases as the mean number of phase singularities  $\langle N_s \rangle_c$ increases.
 We quantify this expectation by plotting $\langle N_s \rangle_{c}$ versus $\langle \xi \rangle_{c}$ in Fig.~\ref{fig:phasesingres} (a); here,
 $\langle \cdot \rangle_c$ denotes an average over several configurations of $V_m$ [see Appendix~\ref{app:Add_mesh_details} for details].
We might argue, \textit{na\"ively}, that we can improve on the mesh-defibrillation scheme by applying $I_{ext}$ directly at the locations of 
phase singularities [e.g., at spiral centers]. This works, indeed, if the system displays only one spiral wave, as shown, e.g., in Refs.~\cite{mulimani2020comparisons,pravdin2020overdrive}. Unfortunately, this control strategy does not work always when there is a spiral-wave turbulence state with many phase singularities~\cite{mulimani2020comparisons}, because this state is sustained by the propagating spiral arms, as we illustrate for the spiral-turbulence patterns in  Fig.~\ref{fig:phasesingres} (b): In the left panel of Fig.~\ref{fig:phasesingres} (b) we show an illustrative pseudocolor plot of $V_m$ for a spiral-turbulence pattern; we apply $I_{ext}$ to the array of electrodes, in the vicinities of phase singularities~\eqref{eq:Ph_In} seen in the middle panel of Fig.~\ref{fig:phasesingres} (b); the right panel of Fig.~\ref{fig:phasesingres} (b) shows that, in spite of the application of this
$I_{ext}$, the spiral-turbulence pattern is sustained because of the propagating spiral arms.

We hypothesize, therefore, and verify in the remaining part of this paper, that, for efficient defibrillation, we should apply $I_{ext}$ along the propagating region of spiral arms. The first step in such a control strategy is the identification of the propagating spiral arms. This can be done most elegantly by monitoring the sodium current $I_{Na}$, which peaks at the propagating front of a spiral arm [see Fig.~\ref{fig:INa_pred} (a) and (b) for an illustrative comparison of $V_{m}$ and $I_{Na}$]
because the sharp peak in the action potential of a cardiac myocyte arises from the depolarizing current associated with the opening of 
voltage-gated sodium ion channels. The plot in Fig.~\ref{fig:phasesingres} (c) of  $\langle\tilde{I}_{Na}^{total}\rangle_{c}$, the mean of the total normalized sodium current~\eqref{eq:Itot}, versus $\langle\xi\rangle_{c}$ shows a distinct trend: $\langle\tilde{I}_{Na}^{total}\rangle_{c}$ increases with decrease in $\langle\xi\rangle_{c}$ because the number of spiral arms increases with $\langle N_{s} \rangle_{c}$ [or, equivalently, as $\langle\xi\rangle_{c}$ decreases]. The application of $I_{ext}$ in these regions is suitable for adjusting the total defibrillation current depending on the spatial correlation length of the spiral-turbulence pattern. We explore this in the next Section~\ref{subsec:U-Netdefib}.

\subsection{U-Net-aided Defibrillation}\label{subsec:U-Netdefib}

We now combine our observations in Section~\ref{subsec:phsing} to develop a U-Net-aided defibrillation scheme. To apply the external current $I_{ext}$ in the vicinity of the spiral arm, whose propagating front is traced by $I_{Na}$, we must have access to this current.
Most experiments measure $V_m$. Therefore, we train a U-Net to extract images of $I_{Na}$ from measured images of $V_m$, which we obtain from our numerical
simulations of the two-dimensional TP06 model~\eqref{eq:TP06}.

We train our U-Net [see Fig.~\ref{fig:INa_pred} (c) and Fig.~\ref{fig:UNet}] to obtain a nonlinear mapping from $\tilde{V}_{m}$ to ${\tilde{I}_{Na}}$, i.e., we normalize our $V_{m}$ and $I_{Na}$ before feeding them into our U-Net [details in Section~\ref{subsec:NN}].
In Fig.~\ref{fig:UNet} of Section~\ref{subsec:NN}, we see that the encoder part of our U-Net has convolutional layers that perform nonlinear operations [for an illustrative expression, see Eq.~\eqref{eq:conv}], and max-pooling layers that achieve dimensionality reduction to obtain compact features. These layers act alternately, capturing abstract representations from the input $\tilde{V}_{m}$, which are then utilized by the decoder part, with convolutional layers and upsampling layers; the latter perform the inverse operations of the max-pooling layer to recover the original spatial dimensions for the associated predictions of $\tilde{V}_{m}$. To enhance the quality of our predictions, we introduce skip connections, where we concatenate the feature maps from the encoder and the decoder parts of our U-Net [see Fig.~\ref{fig:UNet}]. The parameters in our U-Net are trained to minimize the mean-squared-error (MSE) loss~\eqref{eq:loss} between the ground truth $\tilde{I}_{Na}$ and the U-Net prediction.  We normalize the U-Net predictions, explicitly to make the maximum entries of the U-Net to be $1$, and we denote this as $\widehat{\tilde{I}_{Na}}$. For U-Net training, we use values of $\tilde{V}_m$ and $\widehat{\tilde{I}_{Na}}$  defined on $200^{2}$ collocation points
[see Section~\ref{subsec:NN} and Appendix~\ref{app:NNAT} for the details of the U-Net architecture, and the training and testing data]. 

We present $\tilde{I}_{Na}$ and $\widehat{\tilde{I}_{Na}}$ in Fig.~\ref{fig:INa_pred} (d), and to illustrate the quality of our prediction, we also provide $|\tilde{I}_{Na}-\widehat{\tilde{I}_{Na}}|$ [Fig.~\ref{fig:INa_pred} (d)]. Once we obtain  $\widehat{\tilde{I}_{Na}}$ on $200^{2}$ collocation points, we upscale this  to obtain $\widehat{\tilde{I}_{Na}}$ on  $1000^{2}$ collocations points, via bilinear interpolation, which is 
also the number of collocation points we use in our numerical simulations of Eq.~\ref{eq:TP06}.

In Fig.~\ref{fig:INa_pred} (e), we give the MSE loss for the validation data with  various values of the signal-to-noise ratio (SNR) [Eq.~\eqref{eq:SNR} in Section~\ref{sec:SNR}] for our U-Net, trained with and without noise in the input  $\tilde{V}_{m}$. Here, we see that, for high SNR, the U-Net trained with noise performs much better than the one trained without noise. In Fig.~\ref{fig:INa_pred} (f), we give an illustrative pseudocolor plot of $\tilde{V}_{m}$, with SNR=$10$ decibel (dB), and the corresponding predictions $\widehat{\tilde{I}_{Na}}$ from our U-Net, trained both with and without noise.

 \begin{figure*}
    \centering
    \includegraphics[width=\textwidth]{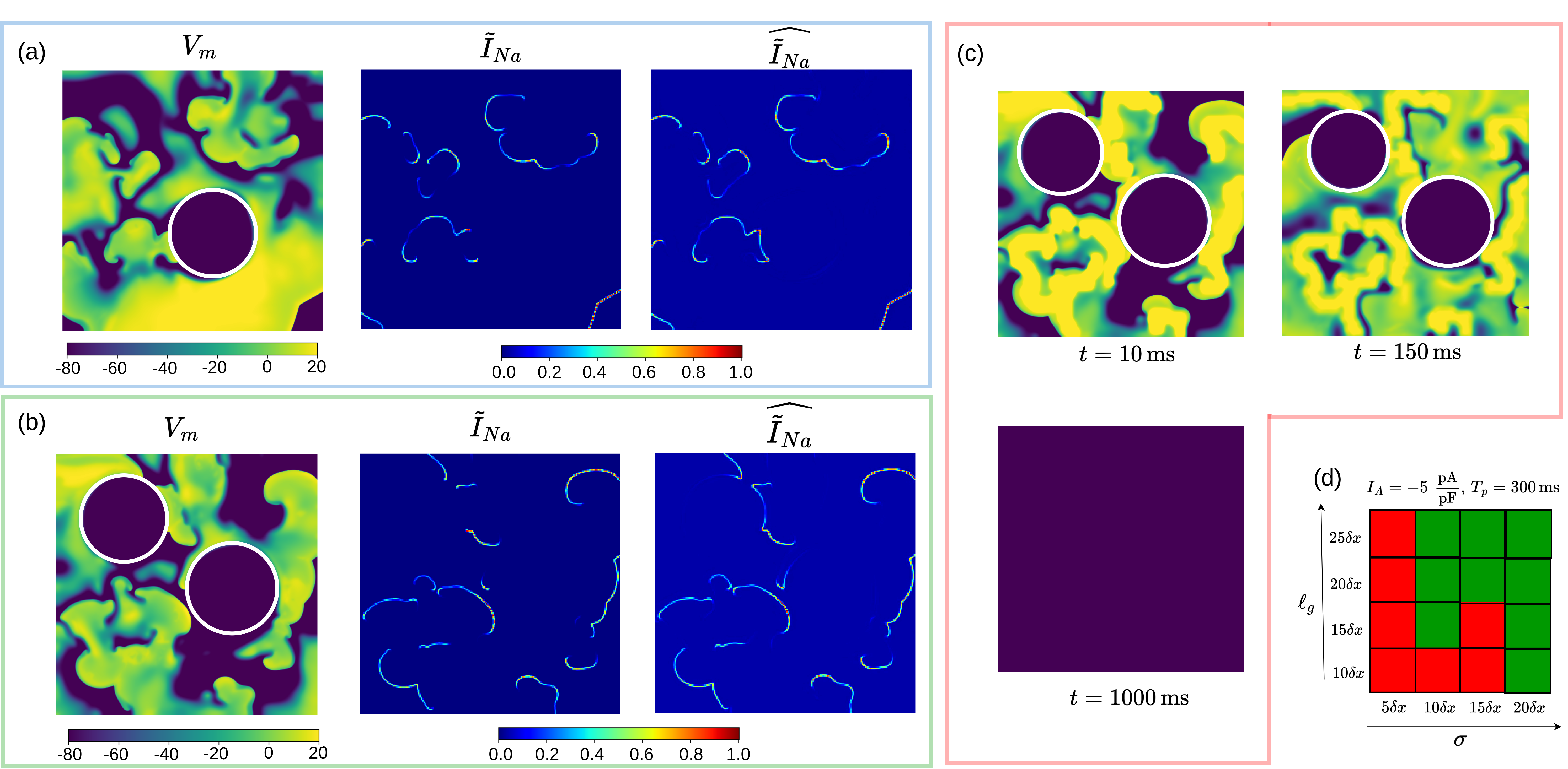}
    \caption{Illustrative pseudo-color plots, for the U-Net predictions ($\widehat{\tilde{I}_{Na}}$) for the $\tilde{I}_{Na}$, when tested with $V_{m}$ with one or two inhomogeneities in (a) and (b), respectively. (c) Elimination of spiral-wave turbulence in (b) [$t=0$]; and in (d) the success of our U-Net-based defibrillation scheme for various choices of $\ell_{\mathrm{g}}$ and $\sigma$, and $I_{A}=-5\,\frac{\textrm{pA}}{\textrm{pF}}$, and $T_{\mathrm{p}}=300$ ms in Eqs.~\eqref{eq:Kernal} and \eqref{eq:Ele}; green and red squares indicate successful and unsuccessful outcomes, respectively.
    }
    \label{fig:inhomo}
\end{figure*}
To eliminate the spiral-turbulence patterns, we start with a lattice of Gaussian electrodes [similar to those in Ref.~\cite{mulimani2020deep}], and then use our U-Net to obtain $\widehat{\tilde{I}_{Na}}$ and then excite these electrodes 
as follows:
\begin{eqnarray}
K(x,y)&=&\Theta[max\{({\widehat{\tilde{I}_{Na}}}(x-l,y-l)-\theta)\,;\nonumber\\
&&l\in[0,2\ell_{g}]\}]\,;\label{eq:Kernal}\\
E(x,y)&=&\sum^{N_{g},N_{g}}_{m,n=1,1}K(x,y) e^{[\frac{-[(x-m\ell_{g})^2+(y-n\ell_{g})^2]}{2\sigma^2}]}\,.
\label{eq:Ele}
\end{eqnarray}
The function $K$ controls the electrodes that are switched on, via the Heaviside function $\Theta$; and $E$ gives the configuration of our Gaussian electrodes, with pulse width $\sigma$, on our square lattice with lattice spacing $\ell_g$;
i.e., we switch on a given electrode at $(m\ell_g,\,n\ell_g)$, if the maximal value of $({\widehat{\tilde{I}_{Na}}}(x-l,y-l,t)-\theta)\geq0$, within a distance of $l=2l_{g}$, along the $x$ and $y$ axes. In Fig.~\ref{fig:INa_pred} (g), we give $E$, for the $\widehat{\tilde{I}_{Na}}$ in Fig.~\ref{fig:INa_pred} (d), for $(\sigma=5\delta x,\,\ell_g=15\delta x)$ and $(\sigma=10\delta x,\ell_{g}=30\delta x)$. For our defibrillation scheme, we indentify the $I_{ext}$~\eqref{eq:TP06} as
\begin{eqnarray}
I_{ext}(x,\,y)&=&I_{A}E(x,\,y),
\label{eq:Iext_E}
\end{eqnarray}
where $I_{A}$ controls the amplitude of the applied current.

Even though our U-Net has been trained with spiral-turbulence patterns, it is able to predict [Fig.~\ref{fig:INa_pred} (h)] $\widehat{\tilde{I}_{Na}}$, for the single spiral wave [Fig.~\ref{fig:INa_pred} (h)]. In Fig.~\ref{fig:INa_pred} (i), we demonstrate that, if we apply $I_{ext}(x,\,y)$ [Eq.~\eqref{eq:Iext_E}] in the vicinity of $\widehat{\tilde{I}_{Na}}$, then the spiral arm collides with the refractory region and the spiral is eliminated.

The pseudocolor plots of $V_{m}$, at $t=10\,\text{ms}$, $t=120\,\text{ms}$,
and $t=1000\,\text{ms}$ in,
respectively, the left, middle, and right sub-panels of Fig.~\ref{fig:defib_result} (a) [$t=0$ in Fig.~\ref{fig:INa_pred} (a)],
illustrate our defibrillation scheme; here, we apply $I_{ext}(x,y)$~\eqref{eq:Iext_E}
for ${T}_p=300\,\text{ms}$, with $I_{A}=-5\frac{\text{pA}}{\text{pF}}$ and the $E(x,y)$ portrayed in the left panel of Fig.~\ref{fig:INa_pred} (g).
Clearly, this defibrillation works in so far as no spiral patterns and excitation waves are visible in the right sub-panel
of Fig.~\ref{fig:defib_result} (a). By contrast, the pseudocolor plots of $V_{m}$ in Fig.~\ref{fig:defib_result} (b) show that the mesh-defibrillation scheme~\ref{subsubsec:MD} does not succeed in removing these spiral waves, even a $t=1000\,\text{ms}$, where we choose $\ell$ such that $2\times \frac{1000}{\ell} \times 3 \delta x \times 1000  \simeq E$ [this is the fraction of the area of the domain covered by the mesh] and apply $I_{ext}=-7.5\,\frac{\text{pA}}{\text{pF}}$ for $300\,\text{ms}$.

The parameters $I_{A},\,T_p,\,\ell_{g},\,\text{and}\,\sigma$ are crucial in our U-Net based defibrillation scheme. In Figs.~\ref{fig:defib_result} (c), (d), and (e), we explore the success of this scheme for various choices of these parameters;
green and red squares indicate successful and unsuccessful outcomes, respectively. 

In Fig.~\ref{fig:defib_result} (f), we plot ${\langle E^{tot} \rangle}_{c}$ [$E^{tot}=\sum^{N_p,N_p}_{x,\,y}E(x,\,y)$; here we use $(\sigma=5\delta x,\ell_{g}=\,15\delta x)$] versus ${\langle \xi \rangle}_{c}$,
where $\langle \cdot \rangle_c$ denotes an average over several configurations of $V_m$ [Appendix~\ref{app:Add_mesh_details}], to show that ${\langle E \rangle}_{c}$ decreases with increasing ${\langle \xi \rangle}_{c}$. This indicates that our 
U-Net-based defibrillation scheme reduces the total applied current as the spatial correlation length  ${\langle \xi \rangle}_{c}$, of the spiral-turbulence pattern, increases.

Finally, in Fig.~\ref{fig:inhomo}, we test our U-Net-based defibrillation scheme in the presence of circular conduction inhomogeneities~\cite{shajahan2007spiral}. In Figs.~\ref{fig:inhomo} (a) and (b), we see that our U-Net predictions for $\tilde{I}_{\mathrm{Na}}$ agree well with the ground truth for illustrative spiral-turbulence patterns in the presence of one and two non-conducting circular inhomogeneities. 
In Fig.~\ref{fig:inhomo} (c), we implement our U-Net-aided defibrillation scheme for the illustrative case of two conduction inhomogeneities [cf. Fig.~\ref{fig:inhomo} (b)]; we use $\ell_{g}=15\delta x$, $\sigma=10\delta x$, $I_{A}=-5\,\frac{\textrm{pA}}{\textrm{pF}}$, and $T_p=300\,\textrm{ms}$. Figure~\ref{fig:inhomo} (d) gives the outcomes of our defibrillation scheme for various values of $\ell_{\mathrm{g}}$ and $\sigma$, and $I_{A}=-5\frac{\textrm{pA}}{\textrm{pF}}$, and $T_{p}=300\,\text{ms}$. Thus, our U-Net-aided defibrillation scheme is successful in eliminating spiral-wave turbulence even if conduction inhomogeneities are present.

\section{Discussion and conclusions\label{sec:dis_con}}

Using our U-Net, we have developed a defibrillation scheme that selects the regions for the application of external current to eliminate spiral-wave turbulence in mathematical models of cardiac tissue. In particular, using the mesh defibrillation scheme, we have shown that spiral-turbulence patterns, with large spatial correlation length ($\xi$), are eliminated more easily than those with small $\xi$. With the voltage pseudocolor plots of the spiral-turbulence patterns, we have used our U-Net-based predictions of sodium currents to predict their activation region to identify the propagating regions of the spiral arms. 
We then apply currents in these regions and develop a novel defibrillation scheme that adjusts the total applied current depending on the spatial correlation of the spiral-turbulence patterns. Our U-Net-aided elimination of spiral-wave turbulence is superior to earlier methods, such as the mesh-defibrillation scheme~\cite{sinha2001defibrillation}, because it requires lesser defibrillation current. Spiral-turbulence patterns in mathematical models for ventricular tissue are the mathematical analogs of life-threatening arrhythmias like ventricular fibrillation. Our study is important, therefore, in the context of targeted-defibrillation strategies, where the regions of application of the defibrillation current are limited to certain areas to reduce the applied energy, e.g., the use of partially insulated defibrillation coils to focus electrical energy on the myocardium~\cite{killu2017novel} and targeting the excitable gap of the reentry~\cite{moreno2022low}. Such strategies are being actively pursued, and it is important to evaluate the efficacy of our defibrillation in such efforts. In particular, detailed investigations, designed to examine defibrillation outcomes via the optimization of our U-Net-based scheme, should be able to advance defibrillation technologies.

\section{Models and Methods\label{sec:method}}

In Section~\ref{subsec:models}, we introduce the models we use for cardiac tissue. In Section~\ref{subsec:NM}, we describe the numerical methods we employ.
The neural networks we employ are described in Section~\ref{subsec:NN}.

\subsection{Models}\label{subsec:models}

We consider the following two models for studying electrical-activation waves in cardiac tissue: the Aliev-Panfilov model~\cite{aliev1996simple}
and the Ten Tusscher-Panfilov (TP06) model~\cite{ten2004model,ten2006alternans} that we describe in Sections ~\ref{subsec:AP} and \ref{subsec:TP06}, respectively. In Section~\ref{subsubsec:SPF}, we define the spatial correlation functions that we require in our study.

\subsubsection{Aliev-Panfilov model}\label{subsec:AP}

The Aliev-Panfilov model~\cite{aliev1996simple,schlemmer2015entropy,mulimani2023overview} consists of two variables: the fast variable $u$, which is a non-dimensionalized transmembrane potential, 
and the slow variable $v$, which accounts for the influence of ion channels. The spatiotemporal evolution of this model is governed by:
\begin{eqnarray}
\frac{\partial u}{\partial t}&=& ku(1-u)(u-a) - uv - I_{ext} + D\nabla^{2}u\,;\nonumber \\
\frac{\partial v}{\partial t}&=&( \epsilon + \frac{m_{1}v}{m_{2}+u})(-v-ku[u-(a+1)])\,;
\label{eq:AP}
\end{eqnarray}
here, $a\,,m_{1},\,m_{2},\,k,\,$ and $ \epsilon$, are model parameters, $D$ is the diffusivity, $\nabla^{2}$ is the spatial Laplacian, and $I_{ext}$ is the external current. In our study, we fix $m_{2}=0.3,\,k=8.0,\,\epsilon=0.01$~\cite{schlemmer2015entropy}, and vary $a,\,m_1$, and $D$. In Section~\ref{sec:results}, we present our results for $a \in [0.07, 0.1]$, $m_{1} \in [0.09, 0.1]$~\cite{schlemmer2015entropy} and $D\in[1,1.5]$.

\subsubsection{Ten Tusscher-Panfilov (TP06) model}\label{subsec:TP06}

The Ten Tusscher-Panfilov 2006 (TP06) model~\cite{ten2004model,ten2006alternans,mulimani2023overview} is an ionically realistic model for the electrical activity of a human ventricular myocite. The spatiotemporal evolution of  this model is given by:
\begin{eqnarray}
\frac{\partial{V_{m}}}{{\partial{t}}}&=& \ D \nabla^{2} V_m - \frac{I_{ion}+I_{ext}}{C_{m}}\,,
\label{eq:TP06}
\end{eqnarray}
where the $V_{m}$ is the transmembrane potential, $C_{m}$ is the membrane capacitance per unit area, and
\begin{eqnarray}
I_{ion} &=& I_{Na}+I_{CaL}+I_{to}+I_{Kr}+ 
I_{Ks}+I_{K1}+I_{NaCa}\nonumber\\&&
+ I_{NaK}+I_{bNa}+I_{bCa}
 +I_{pCa}+I_{pK},
 \label{eq:TP06_1}
\end{eqnarray}
where the total ionic current $I_{ion}$ is the sum of $12$ major ionic currents, which also depend on gating variables [see Appendix~\ref{app:TP06_var}]. We define the fast Na$^+$ current (inward) as
\begin{eqnarray}
I_{Na} &=& {G_{Na}}m^3 hj(V_{m}-E_{Na}), 
\label{eq:INa}
\end{eqnarray}
where $G_{Na}$ is the maximal conductance, $m$, $h$, $j$ are the gating variables, and $E_{Na}$ is the Nernst potential for the fast sodium current ($I_{Na}$) [see Appendix~\ref{app:TP06_var} and Refs.~\cite{ten2004model,ten2006alternans} for the details]. We define the slow delayed rectifier current (outward) as
\begin{eqnarray}
I_{{K}s} &=& G_{Ks}{{x_s}^2}(V_{m}-E_{Ks}),
\label{eq:IKs}
\end{eqnarray} 
where $G_{Ks}$ is the maximal conductance [in our study, we use the base value of $G_{Ks}=0.441$, and rescale by a factor as explained in the Results~\ref{sec:results}], $x_s$ is the gating variable, and $E_{Ks}$ is the Nernst potential for the slow delayed rectifier current ($I_{{K}s}$) [see Appendix~\ref{app:TP06_var} and Refs.~\cite{ten2004model,ten2006alternans}]. In our study, we use the same parameters as in Ref.~\cite{ten2006alternans}, and vary the $G_{Ks}$ in Eq.~\ref{eq:IKs}, and $D$ in Eq.~\ref{eq:TP06} [see Results.~\ref{sec:results}]. 
For the currents $I_{CaL}$, $I_{to}$, $I_{Kr}$, $I_{K1}$, $I_{NaCa}$, $I_{NaK}$, $I_{bNa}$, $I_{bCa}$, $I_{pCa}$
$I_{pK}$ in Eq.~\eqref{eq:TP06_1}, see Refs.~\cite{ten2004model,ten2006alternans}]; here, $I_{ext}$ is the external current. It is useful to calculate the total normalized fast-$Na$ current
\begin{eqnarray}
\tilde{I}^{total}_{Na}&=&\sum_{x,y=1,1}^{N_p,N_p} \tilde{I}_{Na}(x,y),
\label{eq:Itot}
\end{eqnarray}
where $\tilde{I}_{Na}(x,y)$ is obtained by normalizing $I_{Na}(x,y)$, in Eq.~\eqref{eq:INa}, by first taking the absolute value of each entry and dividing by the maximum value in the domain.

\subsubsection{Spatial Correlation Function}\label{subsubsec:SPF}

To quantify the spatial organization of spiral-turbulence patterns that emerge
from our simulations of the PDEs~\eqref{eq:AP} and \eqref{eq:TP06}], we define the spatial correlation function
\begin{eqnarray}
C(R)&=&\frac{\langle\delta \mathcal{V}(\bm r,t)\delta \mathcal{V}(\bm r+\bm{R},t)\rangle}{\langle\delta \mathcal{V}^{2}(\bm{r},t)\rangle}\,; \label{eq:corr} \\
\delta\mathcal{V}(\bm{r},t)&=&\mathcal{V}(\bm{r},t)-\langle\mathcal{V}(\bm{r},t)\rangle;
\end{eqnarray}
where $\mathcal{V}(\bm{r},t)=u(\bm{r},t)$, for the Aliev-Panfilov model~\eqref{eq:AP}, and $\mathcal{V}(\bm{r},t)=V_{m}(\bm{r},t)$, for the TP06 model~\eqref{eq:TP06}, and $R=|\bm R|$. Here, $\langle \cdot \rangle$ denotes the spatial average, over ${\bm{r}}$ and $R=|\bm R|$, at time $t$ [We drop $t$ from our arguments for 
C~\eqref{eq:corr}, as we do not explicitly use its values.].

For the Aliev-Panfilov model~\eqref{eq:AP}, we extract the spatial correlation length $\xi$ from $C(R)$~\ref{eq:corr}, by using the inverse of the slope from the plot of $\log(C(R))$ versus $\log(R)$, in the regions where it is approximately linear. For the TP06 model~\eqref{eq:TP06}, we use a threshold of 0.5 to extract $\xi$ (i.e., $C(\xi)\simeq 0.5$) from $C(R)$ versus $R$.
 
\subsubsection{Signal-to-noise ratio (SNR)\label{sec:SNR}}
From $\tilde{V}_{m}$, we obtain its noisy counterpart $\tilde{V}^{n}_{m}$ with a signal-to-noise ratio, SNR, of a specified decibel (dB) as follows:
\begin{eqnarray}
    \tilde{V}_{m}^{n}(x,\,y)&=&\tilde{V}_{m}(x,\,y)+\delta {V}_{m}^{n}\,,
    \label{eq:SNR}
\end{eqnarray}
where $\delta {V}_{m}^{n}$ is a random variable drawn from $N(0,\,\sigma_{n})$, a Gaussian distribution with mean $0$ and 
standard deviation
\begin{eqnarray*}    
\sigma_{n}&=&\sqrt{\frac{\sum_{x,y}\tilde{V}_{m}^{2}(x,\,y)}{10^{\frac{\text{SNR}}{10}}}}\,.
\end{eqnarray*}

\subsection{Numerical Methods}\label{subsec:NM}
We describe the numerical schemes that we employ to solve the PDEs~\eqref{eq:AP} and \eqref{eq:TP06} in Section~\ref{subsubsec:PDEsolve}.
Section~\ref{subsubsec:IGM} is devoted to the Iyer-Gray method~\cite{iyer2001experimentalist} that is used to track the tips of spiral waves.
In Section~\ref{subsec:NN}, we cover the neural networks that we use.

\subsubsection{Numerical solutions of the PDEs}\label{subsubsec:PDEsolve}

We employ the finite-difference method to evaluate the Laplacian with a five-point stencil in our two-dimensional (2D) simulations; and we use the forward-Euler method for time marching to solve the PDEs~\eqref{eq:AP} and \eqref{eq:TP06}.

For the Aliev–Panfilov model (Section~\ref{subsec:AP}), we use a time step of $\delta t = 0.02$, a spatial grid size of $\delta x = 1$, and $D = 1$~\cite{panfilov2002spiral} [unless stated otherwise], and a domain size of $300\times300$.

For the TP06 model (Section~\ref{subsec:TP06}), we use a time step of $\delta t = 0.02~\text{ms}$, a spatial grid size of $\delta x = 0.025~\text{cm}$, and a membrane capacitance of $C = 1~\mu\text{F}/\text{cm}^2$\cite{ten2006alternans}. The diffusion constant is set to $D = 0.00154\,\text{cm}^2/\text{s}$ [unless stated otherwise], and a domain size of $1000\times1000$.  To update the gating variables in the TP06 model [Section~\ref{subsec:TP06} and Appendix~\ref{app:TP06_var}], we employ the Rush–Larsen scheme~\cite{rush1978practical,mulimani2020comparisons}.
\subsubsection{Spiral-tip tracking: the Iyer-Gray method}\label{subsubsec:IGM}
We use the Iyer-Gray method~\cite{iyer2001experimentalist} to track the spiral tip. We calculate the phase
\begin{eqnarray}
\theta(\bm r,\,t) &=& \tan^{-1}\left( \frac{\mathcal{V}(\bm r,t+\tau)-\langle \mathcal{V}(\bm r,t) \rangle}{\mathcal{V}(\bm r,t)-\langle \mathcal{V}(\bm r,t)\rangle}\right)\,.
\label{eq:IGtheta}
\end{eqnarray}
We then indentify the phase singularities using the condition
\begin{eqnarray}
\oint_{\mathcal{C}} \nabla \theta(\bm{r}) . d \bm{r} = \pm 2 \pi\,,
\label{eq:Ph_In}
\end{eqnarray}
which is satisfied if the contour ${\mathcal{C}}$ encloses one such singularity; in our computations we choose ${\mathcal{C}}$ to be a $3\times3$ square plaquette~\cite{iyer2001experimentalist}.
\subsubsection{Mesh-defibrillation scheme}\label{subsubsec:MD}
\begin{figure}[H]
\centering
\includegraphics[width=0.45\textwidth]{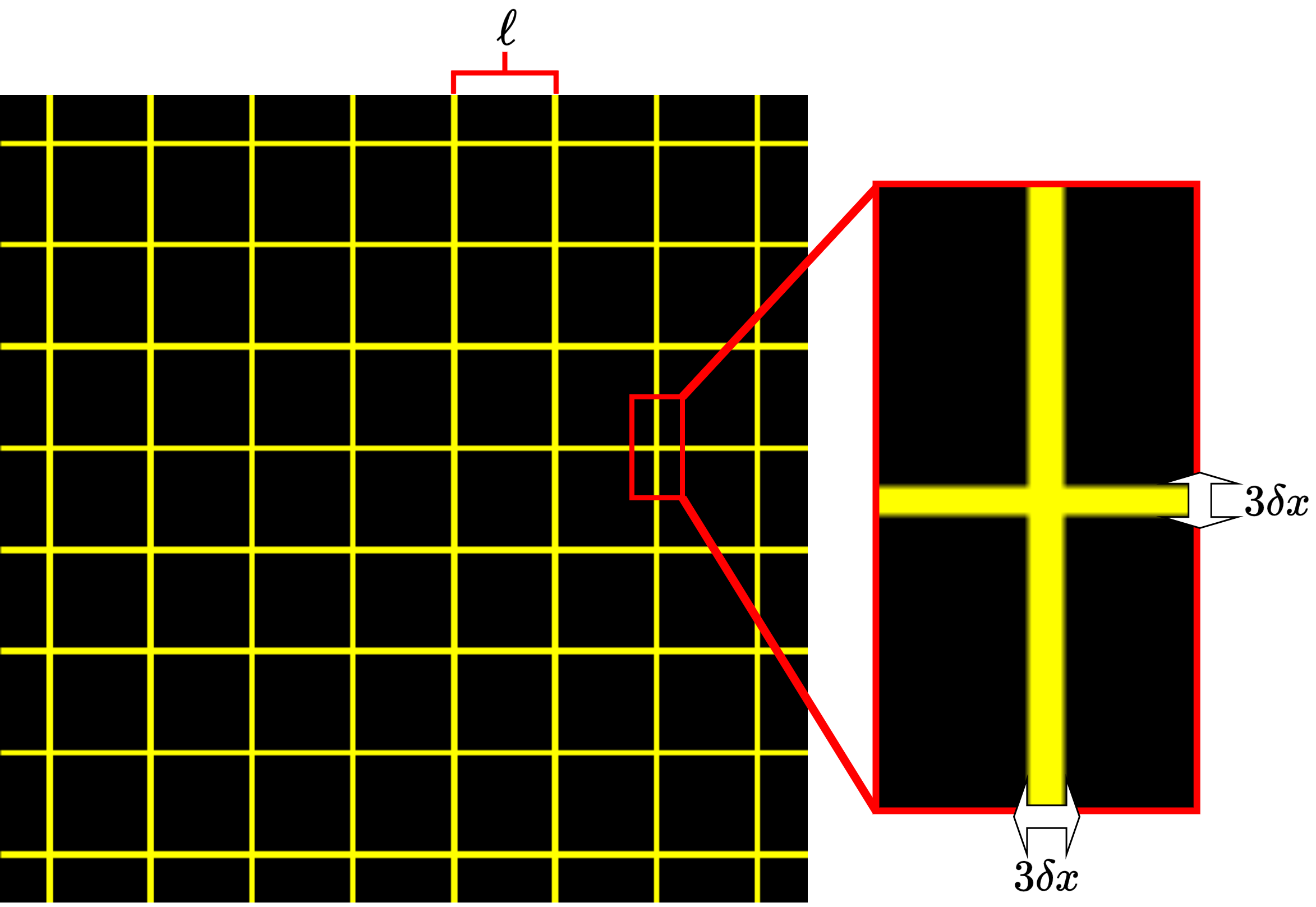}
\caption {Mesh-defibrillation scheme: we apply a current [$I_{ext}$ in Eqs.~\ref{eq:AP} and ~\ref{eq:TP06}] to a square-mesh electrode; the lines that make up this mesh have a width $3\delta x$, and we denote the distance between the lines as $\ell$.}
\label{fig:mesh}
\end{figure}

Our goal is to develop a control (or defibrillation) scheme that eliminates broken spiral waves of electrical activation in the Aliev-Panfilov~\eqref{eq:AP} and TP06~\eqref{eq:TP06} models. To calibrate the effectiveness of our defibrillation scheme, we compare it with the mesh-defibrillation scheme of 
Refs.~\cite{sinha2001defibrillation,shajahan2009spiral}. In the mesh-defibrillation scheme~\cite{sinha2001defibrillation,shajahan2009spiral}, we apply a current [$I_{ext}$ in Eqs.~\ref{eq:AP} and ~\ref{eq:TP06}] to a square-mesh electrode [Fig.~\ref{fig:mesh}]; the lines that make up this mesh have a width $3\delta x$, and the distance between these lines is denoted as $\ell$ in Fig.~\ref{fig:mesh}. We apply a current over a time period, which we denote as $T_p$. The application of this current excites regions near the mesh lines and makes them refractory, so waves of electrical activation get trapped, temporarily, inside the elementary plaquettes that comprise the mesh, and are eventually absorbed at the refractory regions, before they recover and become excitable again. Thus, spiral-wave patterns fail to sustain themselves 
when the control current is applied to this mesh [see Refs.~\cite{sinha2001defibrillation,shajahan2009spiral} for details]. 
\subsection{Neural Networks\label{subsec:NN}}
We begin with the transmembrane potential $V_m(x,y,t)$ that we obtain from our numerical simulation of the TP06 model~\eqref{eq:TP06} [see Section~\ref{subsec:NM}]; the points of our square simulation grid are labeled by the integers $x,y \in [1,1000]$. We then define the normalized transmembrane  potential ${V}^{N}_m(x,y,t) \equiv V_m(x,y,t)/V_{m,max}$, where $V_{m,max}$ is the maximal value of $V_m(x,y,t)$ on this grid. To reduce the size of our training data set, we compute the coarse-grained field $\tilde{V}_m(x,y,t)$ by averaging ${V}^{N}_m(x,y,t)$ over non-overlapping cells with $5 \times 5$ grid points each. We use the field $\tilde{V}_m(x,y,t)$,
defined on $200 \times 200$ grid points, as the input for our U-Net~\cite{ronneberger2015u,babu2025convolutional}. 
The specific U-Net we employ has the architecture shown in the schematic diagram of Fig.~\ref{fig:UNet} [see Table~\ref{tab:UNet} in Appendix~\ref{app} for details]. The encoder part of this U-Net consists of convolutional and max-pooling operations that are applied alternately. The operation performed by our $k^{th}$ convolutional layer [purple arrow in Fig.~\ref{fig:UNet}], to obtain the feature maps
\begin{eqnarray}
F^{C}_k(x,y)&=&\phi\Biggl(\sum_{C',i,j=1,0,0}^{N_f,h-1,w-1}W^{C,C'}_{k,i,j}F^{C'}_{k-1}(x+i,y+j)\nonumber\\
&+&b^{C}_k\Biggr) \label{eq:conv}
\end{eqnarray}
at this layer. The feature maps from the $(k-1)^{\text{th}}$ layer, i.e., $F^{C'}_{k-1}$  are used as inputs to our $k^{th}$ convolutional layer, with filters $W^{C,C'}_{k,i,j}$ and biases $b^{C}_k$; and $\phi$ is a nonlinear activation function [here, we use ReLU]. We use $2 \times 2$ max-pooling filters [red arrows in Fig.~\ref{fig:UNet}], which reduce the number of entries in the feature maps by factors of $2 \, \text{(for the height)}$  and $2 \, \text{(for the width)}$. 

 \begin{figure*}
    \centering
    \includegraphics[width=\textwidth]{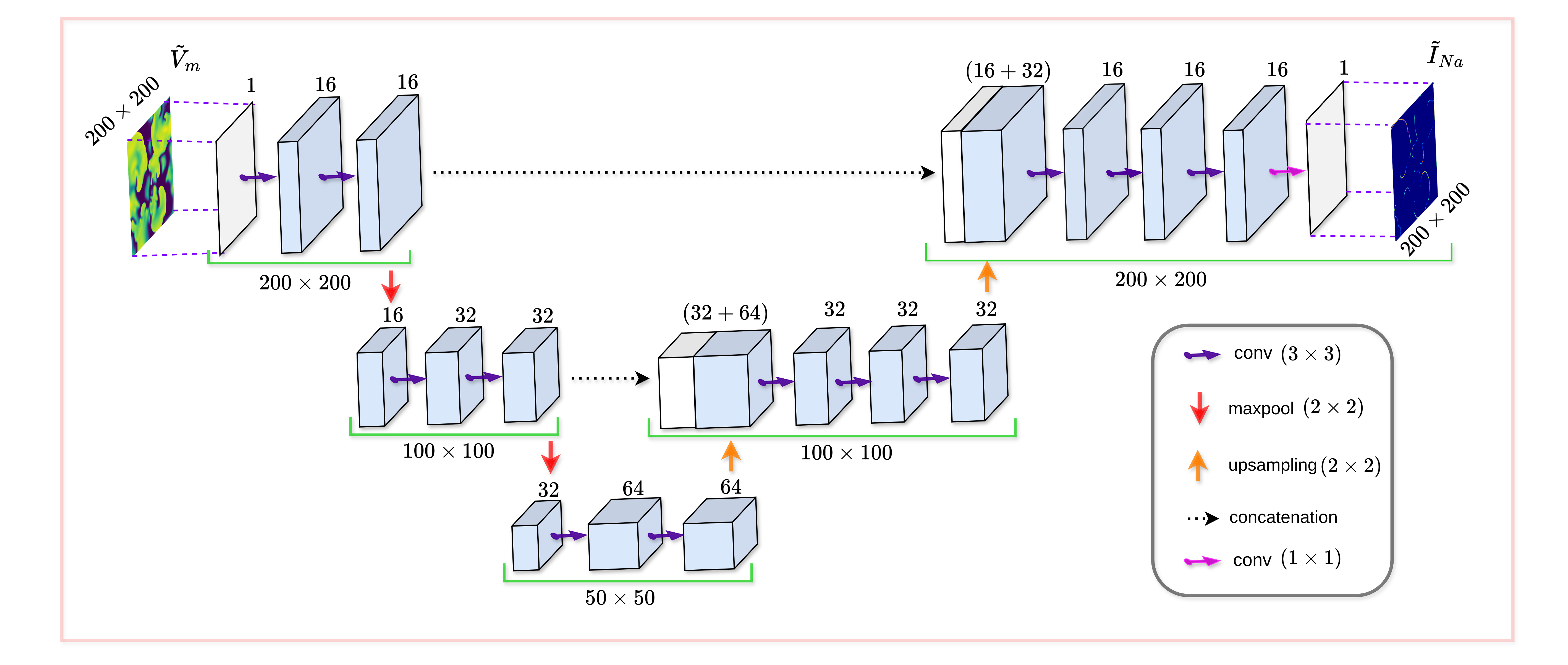}
    \caption{Schematic diagram of our U-Net, which uses $\tilde{V}_{m}$ for training and testing  to predict the $\tilde{I}_{Na}$ in the TP06 model~\eqref{eq:TP06}. The blue slabs indicate layers with feature maps~\eqref{eq:conv}; the numbers at the bottom of the slab denote the width and height of the feature maps in the layer; and the numbers at the top indicate the number of such feature maps in the layer. The purple, red, and orange arrows represent \textit{convolution, maxpooling,} and \textit{upsampling} operations, respectively. The dotted arrows represent the \textit{concantenations} required for the \textit{skip connections} in our U-Net.}
    \label{fig:UNet}
\end{figure*}
The decoder of our U-Net is similar to the encoder, but with upsampling layers [orange arrows in Fig.~\ref{fig:UNet}]. We add padding to recover the original spatial dimensions. We introduce skip connections in our U-Net via \textit{concatenation} [dotted arrows in Fig.~\ref{fig:UNet}]; here, we concatenate the feature maps from the encoder of our U-Net with those of the decoder. These skip connections are known to capture fine-grained details [see, e.g., Ref.~\cite{ronneberger2015u}].

With $\tilde{V}_{m}(x,y,t)$ as the input, we train our U-Net to predict $\tilde{I}_{Na}(x,y,t)$, the normalized [we consider the magnitude of entries for ${I}_{Na}(x,y,t)$] and coarse-grained sodium current obtained from $I_{Na}(x,y,t)$ in Eq.~\eqref{eq:INa}, with  the loss function
\begin{equation}
\text{MSE}=\left\langle\frac{1}{N^{2}_{p}}\sum_{x,y=1,1}^{N_p,N_p} \left[{\widehat{\tilde{I}_{Na}}}(x,y,t) - \tilde{I}_{Na}(x,y,t)\right]^ {2}\right\rangle,
\label{eq:loss}
\end{equation}
where $\widehat{\tilde{I}_{Na}}(x,y,t)$ is our neural-network prediction, ${N_p}^{2}$ is the number of points in the domain, and  $\left\langle\cdot\right\rangle$ denotes the average over the training data set.

\begin{acknowledgments}  
\hspace{10pt}We thank the  the Anusandhan National Research Foundation (ANRF), the Science and Engineering Research Board (SERB), and the National Supercomputing Mission (NSM), India, for support,  and the Supercomputer Education and Research Centre (IISc), for computational resources.
\end{acknowledgments}

\section*{Data and code availibility}
\hspace{10pt}The data and code utilized in this study can be made available from the authors upon reasonable request.

\section{Appendices}
\label{app}

\subsection{Additional simulation details\label{subsec:sim_det}}

\par{\textit{Aliev-Panfilov model:}} We generate spiral-wave turbulence in the Aliev-Panfilov model~\eqref{eq:AP} as follows: we start with a broken-wavefront initial condition~\cite{shajahan2007spiral}, which evolves into a spiral wave and subsequently breaks into a spiral-wave turbulence in the parameter range given in Sections~\ref{subsec:ReAP} and ~\ref{subsec:AP}. In Section~\ref{subsec:ReAP} and Fig.~\ref{fig:AVTP06_supp}, for each set of parameters, we discard the first $10^5$ time steps. Then we consider $20$ pseudocolor plots of $V_m$ (snapshots),  spaced at intervals of $5000$ steps, from the next $10^5$ time iterations; the average $\langle\cdot\rangle_{s}$ is taken over these $20$ snapshots.

\par{\textit{TP06 model:}} We generate spiral-wave turbulence in the TP06 model~\eqref{eq:TP06} via the S1-S2 cross-field protocol~\cite{nayak2014spiral,hajian2024weak}: We inject the S1 plane wave at one edge of the simulation domain; once its tail reaches the centre of the domain, we apply an additional stimulus (S2) perpendicular to the S1 plane wave; this leads to a spiral wave that breaks to yield spiral turbulence [in the parameter range described in Sections~\ref{subsec:ReTP06} and ~\ref{subsec:TP06}]. In Section~\ref{subsec:ReTP06} and Fig.~\ref{fig:AVTP06_supp}, for each set of parameters, we discard the first $250000 \delta t$  ($5000\,\text{ms}$) time steps. Then we consider $25$ pseudocolor plots of $V_m$ (snapshots),  spaced at intervals of $10000 \delta t$  ($200\,\text{ms}$), from the next  $250000\delta t$; the average $\langle\cdot\rangle_{c}$ is taken over these $25$ snapshots.

We calculate $\xi$ and $N_s$ [see Eq.~\eqref{eq:Ph_In}] for 100 snapshots [50 snapshots spaced at intervals of $100\,\text{ms}$ (see above) for each parameter value in Fig.~\ref{fig:AV_TP06_tableST} (i)]. We then calculate $\langle N_p \rangle_c$ and $\langle \xi \rangle_c$, where the average $\langle \cdot \rangle_c$ is performed over the $V_m$ configurations selected in bins. We identify the individual bins as $40 \delta x + i * 15 \delta x \leq \xi < 40 \delta x + (i+1) * 15 \delta x$, where $i \in [0, 1, 2, 3]$.

\subsection{Additional details of mesh defibrillation outcomes\label{app:Add_mesh_details}}

\begin{figure*}
    \centering
    \includegraphics[width=\textwidth]{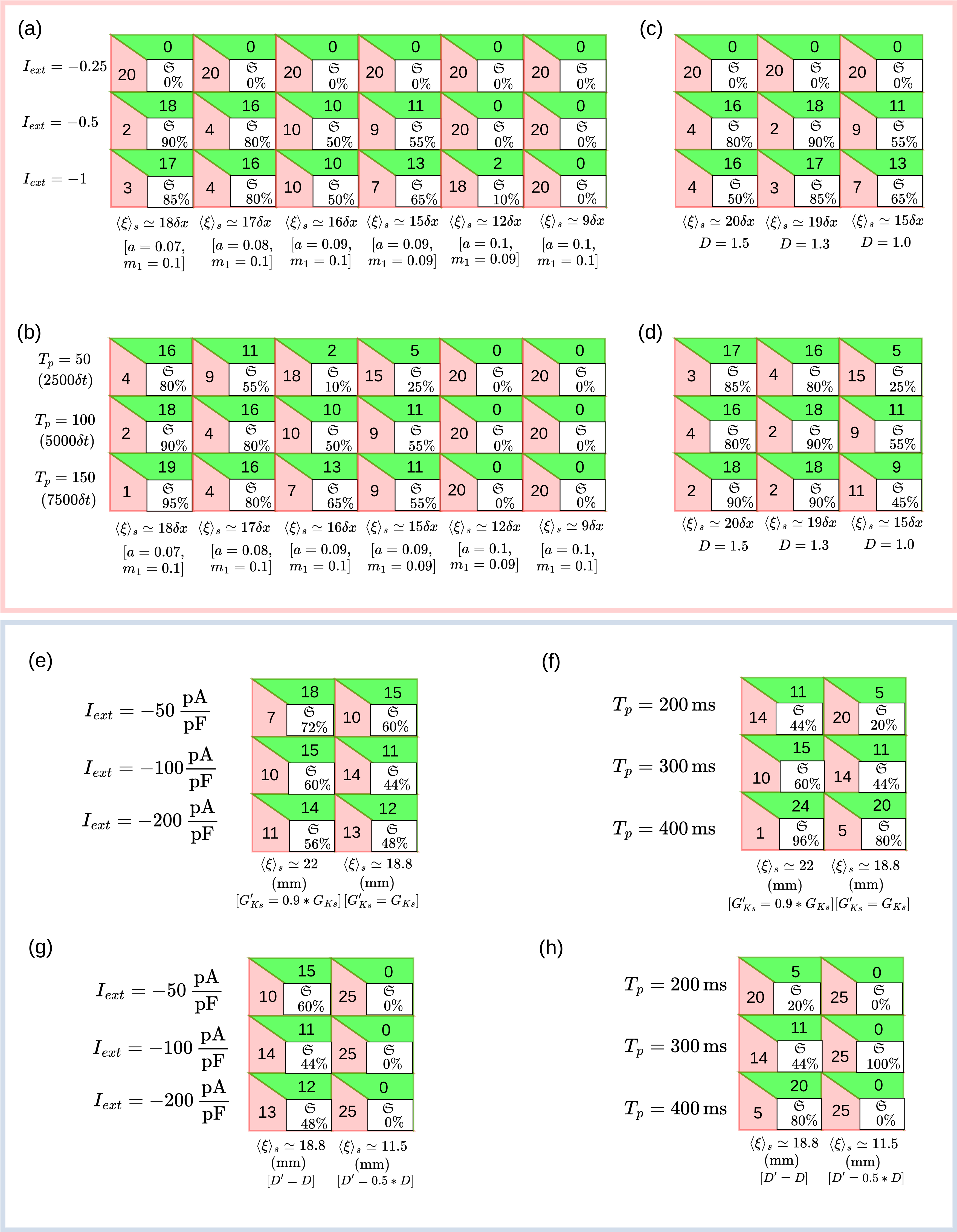}
    \caption{(a), (b), (c), (d): The upper green quadrilateral (lower red quadrilateral) gives the number of successful (unsuccessful) cases of the elimination of spiral-turbulence patterns, via mesh defibrillation [Section ~\ref{subsubsec:MD}] for $20$ spiral-turbulence patterns for each set of parameters, $a$ and $m_{1}$ [with all other parameters fixed in Eq.~\eqref{eq:AP}], which are given at the bottom, along with the mean correlation length $\langle\xi\rangle_{c}$ [see text]; $\mathfrak{S}$ is the elimination success rate.  (e), (f), (g), (h) are the TP06~\eqref{eq:TP06} counterparts of (a), (b), (c), (d) respectively [see text].}
    \label{fig:AVTP06_supp}
\end{figure*}

In Figs.~\ref{fig:AVTP06_supp} (a) and (b), we give the outcomes of the mesh-defibrillation scheme~\ref{subsubsec:MD} for the Aliev-Panfilov model~\eqref{eq:AP}, with varying $I_{ext}$ [with $\ell=30\delta x$ and $T_p=100\,(5000\delta t)$], and $T_p$ [with $\ell=30\delta x$ and $I_{ext}=-0.5$] respectively, for different values of parameters $a$ and $m_{1}$, keeping others fixed in Eq.~\eqref{eq:AP} [see Section.~\ref{sec:method} for a full description]. Figs.~\ref{fig:AVTP06_supp} (c) and (d) are similar to (a) and (b), but with varying $D$, keeping other parameters fixed in Eq.~\eqref{eq:AP}.

In Figs.~\ref{fig:AVTP06_supp} (e) and (f), we give the outcomes of the mesh-defibrillation scheme~\ref{subsubsec:MD} for the TP06 model~\eqref{eq:TP06}, with varying $I_{ext}$ [with $\ell=200\delta x\,(50\,\text{mm})$ and $T_p=15000\delta t\,(300\,\text{ms})$] and $T_p$ [with $\ell=200\delta x\,(50\,\text{mm})$ and $I_{ext}=-100\,\frac{\text{pA}}{\text{pF}}$], respectively, for different values of $G_{Ks}$, keeping other parameters fixed in Eq.~\eqref{eq:TP06} [see Section.~\ref{sec:method} for a full description]. Figures~\ref{fig:AVTP06_supp} (g) and (h) are similar to (e) and (f), but with varying $D$, keeping other parameters fixed in Eq.~\eqref{eq:TP06}.

\subsection{\label{app:NNAT} Data-sets and neural-network training}

To train our U-Net to predict the $\tilde{I}_{Na}$ from $\tilde{V}_{m}$ [Fig.~\ref{fig:INa_pred}], we obtain the training- and the validation-data sets as follows: we use the  parameters $G_{Ks}=0.441$ and $D = 0.00154\,\frac{\text{cm}^2}{\text{s}}$ in the TP06 model [all other parameters have their standard values~\cite{ten2006alternans}].  We discard data from the first $250000\delta t$ ($5000\,\text{ms}$) times. We then obtain $750$ $\tilde{V}_{m}$ and $\tilde{I}_{Na}$ snapshots from the next $750000\delta t$ ($15000\,\text{ms}$). From these $750$
snapshots, we use $500$ configurations of $\tilde{V}_{m}$ and $\tilde{I}_{Na}$ for the training data; and we use the remaining $250$ configurations of $\tilde{V}_{m}$ and $\tilde{I}_{Na}$ for the validation data set [Fig.~\ref{fig:INa_pred} (e)]. While training the U-Net with noise, we prepare our training data set as follows: in addition to the $500$ configurations, we randomly select sets of $125$ configurations out of the above $500$, with noise strength varying from SNR=$5$ dB to $35$ dB. 

We give the architecture of our U-Net to map the $200^{2}$ $\tilde{V}_{m}$ to $\tilde{I}_{Na}$ in the Table.~\ref{tab:UNet}. We train our U-Net with the Adam optimizer~\cite{kingma2014adam}, with an initial learning rate of $10^{-3}$, for $200$ epochs. We use a batch size of $64$.

\begin{table}[htbp]
    \centering
    \renewcommand{\arraystretch}{1}
    \begin{tabular}{|c|c|c|c|}
       \hline
        Layer &  Type & Details & Parameters  \\
        \hline
          &  Input & $\tilde{V}_{m}$ on $200^2$  & 0 \\
          &   & collocation points &  \\
        \hline
        1 &  Conv2D & $3^{2}$ filter + 16 channels  & 160 \\
         &           &  + ReLU Activation &  \\
        \hline
        2 &  Conv2D & $3^{2}$ filter + 16 channels  & 2320 \\
         &           &  + ReLU Activation &  \\
        \hline
        3 &  Maxpool2D & $2^{2}$ filter &  0  \\      
        \hline
        4 &  Conv2D & $3^{2}$ filter + 32 channels  & 4640 \\
         &           &  + ReLU Activation &  \\
        \hline
        5 &  Conv2D & $3^{2}$ filter + 32 channels  & 9248 \\
         &           &  + ReLU Activation &  \\
        \hline
        6 &  Maxpool2D & $2^{2}$ filter &  0  \\     
        \hline
        7 &  Conv2D & $3^{2}$ filter + 64 channels  & 18496 \\
         &           &  + ReLU Activation &  \\
        \hline
        8 &   Conv2D & $3^{2}$ filter + 64 channels  & 36928 \\
         &           &  + ReLU Activation &  \\
        \hline  
        9 &  Upsample2D & $2^{2}$ filter &  0  \\    \hline
        10 &  Concatenate & Layer 5 + Layer 9 &  0  \\
        \hline
        11 &   Conv2D & $3^{2}$ filter + 32 channels  & 27680 \\
         &           &  + ReLU Activation &  \\
        \hline
        12 &  Conv2D & $3^{2}$ filter + 32 channels  & 9248 \\
         &           &  + ReLU Activation &  \\
        \hline
        13 &  Conv2D & $3^{2}$ filter + 32 channels  & 9248 \\
         &           &  + ReLU Activation &  \\
        \hline
        14 &  Upsample2D & $2^{2}$ filter &  0  \\
        \hline
        15 &  Concatenate & Layer 2 + Layer 13 &  0  \\
        \hline
        16 &  Conv2D & $3^{2}$ filter + 16 channels  & 6928 \\
         &           &  + ReLU Activation &  \\
        \hline
        17 &  Conv2D & $3^{2}$ filter + 16 channels  & 2320 \\
         &           &  + ReLU Activation &  \\
        \hline
        18 &  Conv2D & $3^{2}$ filter + 16 channels  & 2320 \\
         &           &  + ReLU Activation &  \\
        \hline
        19 &  Conv2D & $1^{2}$ filter + 1 channel  & 17 \\
         &           &  + Linear Activation &  \\
        \hline             
          &  Output & $\tilde{I}_{Na}$ on $200^2$  & 0 \\
          &   & collocation points &  \\
          \hline
    \end{tabular}
    \caption{The 2D encoder-decoder CNN, which we use to map the $200^{2}$  $\tilde{V}_{m}$ to the $200^{2}$ $\tilde{I}_{Na}$. We give the layer numbers (column 1), their types (column 2), their details (column 3), 
    and the parameters (column 4).}
    \label{tab:UNet}
\end{table}

\subsection{\label{app:TP06_var} Gating variables}
The currents in Eqs.~\eqref{eq:INa} and ~\eqref{eq:IKs} are voltage gated, and the gating variables, denoted generically by $g$, obey ODEs of the form
\begin{eqnarray*}
\frac{dg}{dt} & = & \frac{g_{\infty}-g}{\tau_{g}},
\end{eqnarray*}
where $g_{\infty}=\frac{\alpha_{g}}{(\alpha_{g}+\beta_{g})}$ is the
steady-state value of the gating variable $g$, and $\tau_{g}=\frac{1}{\alpha_{g}+\beta_{g}}$
is its time constant. Here, $\alpha_{g}$ and $\beta_{g}$ are
voltage-dependent rate constants associated with the gate $g$; e.g., for $m$ in Eq.~\eqref{eq:INa}, we have
\begin{eqnarray}
m_{\infty} &=& \frac{1}{\left[1+\exp(\frac{-56.86-V}{9.03})\right]^2},\\\nonumber
\alpha_m &=& \frac{1}{1+\exp(\frac{-60-V}{5})},\\\nonumber
\beta_m &=& \frac{0.1}{1+\exp(\frac{V+35}{5})}+\frac{0.1}{1+\exp(\frac{V-50}{200})}.
\end{eqnarray}
For the full details of all the currents and gating variables for the TP06 model~\ref{subsec:TP06} see Refs.~\cite{ten2004model,ten2006alternans}.
\newpage
\bibliography{references}

\end{document}